\documentclass[lettersize,journal]{IEEEtran}
\usepackage{amsmath,amsfonts}
\usepackage{algorithmic}
\usepackage{algorithm}
\usepackage{array}
\usepackage{textcomp}
\usepackage{stfloats}
\usepackage{url}
\usepackage{verbatim}
\usepackage{graphicx}
\usepackage{cite}
\usepackage{multicol}
\usepackage{amssymb}
\usepackage{color}
\usepackage{lipsum}
\usepackage{mathtools}
\usepackage{mathtools, nccmath}
\usepackage{cuted}
\usepackage[short]{optidef}
\usepackage{comment}
\usepackage[utf8]{inputenc}
\usepackage{subcaption}
\usepackage{csquotes}
\usepackage{epstopdf}

\usepackage{algorithm}
\usepackage{algorithmic}
\begin{document}

	%\title{A Novel Constellation Rotation-based Two-stage LLR Detector for OFDM-IM NOMA}
	\title{A Novel Rotated Constellation-based LLR Detector for Flexible NOMA based OFDM-IM Scheme}
	
\author{Subham Sabud,~\IEEEmembership{Graduate Student Member,~IEEE}, and Preetam Kumar,~\IEEEmembership{Senior Member,~IEEE}
	\thanks{This work is funded by the Ministry of Electronics and Information technology (MeitY), Government of India, Project no: 0429. (\textit{Corresponding author: Subham Sabud})}
	\thanks{The authors are with the Department of Electrical Engineering, Indian Institute of Technology Patna, Bihar, India-801106 (emails: subham$\_$2011ee15@iitp.ac.in, pkumar@iitp.ac.in).}}
	
        \maketitle
	\begin{abstract}
	OFDM-IM NOMA is a newly created flexible scheme for future generation communication systems. For the downlink OFDM-IM NOMA system, a low-complexity \enquote{rotated constellation based log likelihood ratio (LLR) detector} has been proposed in this work. This detector is able to significantly reduce the complexity by employing the rotating constellation-based concept and the log-likelihood ratio-based algorithm together. Complexity analysis and simulation results show that the proposed detector achieves significantly lower computational complexity and much better error performance than the earlier introduced detectors under different scenarios for the OFDM-IM NOMA scheme. 
	\end{abstract}
	\section{Introduction}
	Non-orthogonal multiple access (NOMA) has established itself as a potential candidate for addressing the high connectivity density environment and massive data load in 5G, 6G, and beyond wireless communication networks. In power-domain NOMA \cite{8114722}, the same time and frequency resources are shared by several users, who are separated by various power levels according to their channel condition. NOMA can provide high spectral efficiency, low latency, massive connectivity and fairness.
	\par On the other hand, orthogonal frequency division multiplexing (OFDM) is a popular multicarrier transmission technique that has the ability to successfully counter the inter-symbol interference (ISI) brought on by a frequency-selective channel. Using the spatial modulation (SM) concept on OFDM, different techniques have been proposed in \cite{5449882,6162549,6587554}. Among them, index modulated OFDM (OFDM-IM) \cite{6587554} has proven to be the most promising one. The M-ary signal constellation, as well as the subcarrier indices, carry information in OFDM-IM, Unlike the classical OFDM. As a result, OFDM-IM has better spectral efficiency, error performance, and energy efficiency than the traditional OFDM. Using the IM and SM paradigm in conjunction with cooperative NOMA, several excellent studies have been published. An OFDM-IM system built on co-operative NOMA (C-NOMA) is suggested in \cite{9078069} and is known as CIM-NOMA. The CDM-NOMA technique, which is basically a co-operative NOMA based dual mode OFDM-IM scheme with superior error performance than the generalised CIM-NOMA (GCIM-NOMA), is introduced in \cite{9082642}.\cite{9142447} illustrates the use of a novel technology named C-NOMA-GQSM in multi-vehicle networks. A new energy-efficient, spectrally-efficient, and flexible transmission method called OFDM-IM NOMA \cite{9140042} has recently been proposed by merging the OFDM-IM and NOMA techniques, where OFDM-IM provides flexibility with a tunable subcarrier activation ratio and NOMA provides flexibility with a tunable power allocation coefficient.
	 \cite{9324769} discusses a rather similar technique called IM-NOMA, which differs from the previously proposed NOMA-MCIK \cite{8011316} and Hybrid IM-NOMA \cite{9027834} schemes. %OFDM-IM NOMA is a more energy efficient, spectral efficient, and flexible scheme compared to the conventional NOMA-OFDM \cite{TRIVEDI20199}.
	   In conventional NOMA-OFDM, all the available subcarriers are in use for both the users, whereas, in the OFDM-IM NOMA, we have the flexibility of activating the subcarriers partially based on the user's needs. In spite of having all the benefits listed above, the OFDM-IM NOMA \cite{9140042} system's ML-based detection approach suffers from very high computational complexity. In \cite{9003390}, a low-complexity LLR-based detection algorithm is suggested, which is applicable to a more generalized IM-MA scheme. However, in \cite{9348437} the IM-NOMA \cite{9324769} system uses a constellation rotation-based low-complexity SIC detection technique and recently for the OFDM-IM NOMA \cite{9140042} system, a low complexity two-stage LLR detector has been introduced in \cite{9844730}.
	\par In this letter, a novel \enquote{rotated constellation based LLR detector} is proposed to address the issue of high detection complexity in the downlink OFDM-IM  based flexible  NOMA scheme. This proposed detector substantially reduces the complexity by combining the constellation rotation-based concept and the log-likelihood ratio-based algorithm together. Additionally, this detector provides an improved error performance. In this system model, User A's and User B's symbols are taken from a $\pi/2$-rotated and a $0$-rotated constellation, respectively. Therefore, there will not be any interference between user A's and user B's transmitted signals. As a result, both user A and user B can directly decode their own signals from the superimposed signal without performing the successive interference cancellation (SIC) process. On the other hand, at the receiver side, the purpose of determining the first stage log-likelihood ratio is to identify the active indices, while the purpose of the second stage log-likelihood ratio is to decode the symbols corresponding to the identified active indices.
	
	\section{OFDM-IM NOMA WITH ROTATED CONSTELLATION BASED ML DETECTOR}
	This section explains the transceiver architecture of downlink OFDM-IM NOMA employing the rotated constellation based ML detector.
	A total of $m_{\beta}$ information bits pass through the transmitter for user $\beta$. Here $\beta \in$ $\{A, B\}$, where user A is the near user and user B is the far user.  $m_{\beta}$ bits are split into $G_{\beta}$ groups, each of which carries $p_{\beta}$ bits, so  $p_{\beta}=m_{\beta}/G_{\beta}$.
	In order to build an active indices set  $I_{\beta}(g) =$ $\left\{i_{\beta, 1}(g), \ldots, i_{\beta, k_{\beta}}(g)\right\}$ for the $g^{th}$ subblock and user $\beta$, the first $p_{\beta, 1}$ bits among the  $p_{\beta}$ bits utilize a look-up table approach \cite{6587554} to choose the $k_{\beta}$ active indices from the $n_{\beta}$ available indices. The remaining $p_{\beta, 2}$ bits are mapped to the $k_{\beta}$ active indices' corresponding $M_{\beta}$-ary modulated symbols. Here $g= 1,\ldots, G_{\beta}$.
	In this constellation rotation-based system model, the near user (A) and the far user (B) will take symbols from a $\pi/2$- rotated constellation and a $0$- rotated constellation, respectively. For $g^{th}$ subblock of user A, the vector of modulated symbols is presented as $\boldsymbol{s}_{A}(g)=\left[S_{A, 1} (g), \ldots,S_{A, k_{A}}(g)\right]$, where $S_{ A, \tau}(g) \in e^{j\pi/2}\hspace {1mm} Q_{M_{A}}={Q}'_{M_{A}}$ and $\tau=1, \ldots, k_{A}$. For $g^{th}$ subblock of user B, the vector of modulated symbols is presented as $\boldsymbol{s}_{B}(g)=\left[S_{B, 1} (g), \ldots,S_{B, k_{B}}(g)\right]$, where $S_{ B, \iota }(g) \in  Q_{M_{B}}={Q}'_{M_{B}}$ and $\iota =1, \ldots, k_{B}$. Here ${Q}'_{M_{A}}=\{q_{0}^{A}, \ldots, q_{M_{A}-1}^{A}\}$ and ${Q}'_{M_{B}}=\{q_{0}^{B}, \ldots, q_{M_{B}-1}^{B}\}$ represent the $\pi/2$- rotated constellation corresponding to user A and the $0$- rotated constellation corresponding to user B, respectively. Therefore, each subblock fetches a total of $p_{\beta}=p_{\beta, 1}+p_{\beta, 2}=\left\lfloor\log _{2}\left(C\left(n_{\beta}, k_{\beta}\right)\right)\right\rfloor+k_{\beta} \log _{2}\left(M_{\beta}\right)$ bits of information. The main OFDM-IM block for user $\beta$ is produced as $\boldsymbol{x}_{\beta}=[x_{\beta}(1) \ldots x_{\beta}(N)]^T\in\mathbb{C}^{N\times1}$, taking into account $I_{\beta}(g)$ and $\boldsymbol{s}_{\beta}(g)$ for all the subblocks. Here $N=n_{\beta} G_{\beta}$ represents the overall number of subcarriers. The superimposed signal in the frequency domain is obtained as $\boldsymbol{x}_{sc}=\sqrt{\gamma P_{T}}\boldsymbol{x}_{A}+\sqrt{(1-\gamma) P_{T}}\boldsymbol{x}_{B}$, where $\gamma$ and $P_{T}$ are the power allocation factor and the total transmit power per subcarrier, respectively. Per subcarrier, the average power assigned to user A and user B is $P_{A}=\gamma P_{T}$ and $P_{B}=(1-\gamma) P_{T}$, respectively. After applying the inverse fast fourier transform (IFFT) and cyclic prefix addition (CP) operation on $\boldsymbol{x}_{sc}$, the resultant signal is transmitted through a frequency-selective Rayleigh fading channel. The time domain portrayal of the channel impulse response coefficient vector for user $\beta$ is $\boldsymbol{h}_{T, \beta}=\left[h_{T, \beta}(1) \ldots h_{T, \beta}(L)\right]^{T}$, where $h_{T, \beta}(\rho), \rho=1 \ldots L$ are circularly symmetric and obey ${\cal CN}\left(0, \frac{1}{L}\right)$ distribution.
	\par Following the CP removal and fast fourier transform (FFT) operations at the receiver side, the signal received for user $\beta$ in the frequency domain can be represented as $\boldsymbol{y}_{\beta}=\boldsymbol{x}_{sc} \operatorname{diag}\left(\boldsymbol{h}_{F, \beta}\right)+\boldsymbol{w}_{ \beta}=\left[y_{\beta}(1) \ldots y_{\beta}(N)\right]^{T} $. Here, $\boldsymbol{h}_{F, \beta}=$ FFT $\left\{\boldsymbol{h}_{T, \beta}\right\}=\left[h_{F, \beta}(1) \ldots h_{F, \beta}(N)\right]^{T}$ is the channel vector in the frequency domain following ${\cal CN}\left(0, \varrho_{\beta}^{2}\right)$ distribution and $\boldsymbol{w}_{\beta}=\left[w_{\beta}(1) \ldots w_{\beta}(N)\right]^{T}$ is the noise vector that obeys ${\cal CN}\left(0, \sigma_{N_{F}}^{2}\right)$ distribution. Here $\varrho_{\beta}^{2}$ represents the channel gain for user $\beta$, and due to the distance factor it is assumed that $\varrho_{A}^{2} \geq \varrho_{B}^{2}$. As a result, $P_{B}>P_{A}$. Due to the constellation rotation based system model, the transmitted signal of user A and user B will not interfere with each other. Therefore, both user A and user B can directly decode their own signal from the superimposed signal. $E_{\beta}=\left\{\boldsymbol{e}_{\beta, 1}, \boldsymbol{e}_{\beta, 2}, \ldots , \boldsymbol{e}_{\beta,2^{p_{\beta}}}\right\}$ represents the set containing all feasible subblock realizations for performing the rotated constellation based ML detection for user $\beta$.\\
	The $g^{th}$ subblock of the user $\beta$'s signal is directly decoded using the \enquote{rotated constellation based ML detector} as follows. 
	\begin{equation}
		\widehat{\boldsymbol{x}}_{\beta}^{g}=\underset{\boldsymbol{e}_{\beta} \in E_{\beta}}{\arg \: \min}\left\|\boldsymbol{y}_{\beta}^{g}-\sqrt{P_{\beta}} diag(\boldsymbol{h}_{F,\beta}^{g})\boldsymbol{e}_{\beta}\right\|^{2}
		\label{rotated ML},
	\end{equation}
	where
	$\boldsymbol{h}_{F,\beta}^{g}=\left[h_{F,\beta}\left(n_{\beta}(g-1)+1\right) \ldots h_{F,\beta}\left(n_{\beta} g\right)\right]^{T} \in \mathbb{C}^{n_{\beta} \times 1}$, and $\boldsymbol{y}_{\beta}^{g}=\ [{y}_{\beta}\left (n_{\beta}\left ( g -1 \right )+1  \right )\ldots$ $ y_{\beta}\left (n_{\beta}\hspace{0.5mm}g\right )]^{T}$ $\in\mathbb{C}^{n_{\beta} \times 1}$ are the channel vector and the received signal vector for the $g^{th}$ subblock of user $\beta$, respectively.
	Here $\beta \in$ $\{A, B\}$.
	
 	\section{PROPOSED DETECTOR}
	In this section, the proposed \enquote{rotated constellation-based LLR detector} has been discussed.\\
	In this system model, the symbols of user A and user B are taken from a $\pi/2$- rotated and a $0$- rotated constellation, respectively. Therefore, user A's and user B's transmitted signals will not interfere with each other. As a result, both user A and user B can decode directly their signals from the superimposed signal.
	%Employing a \enquote{rotated constellation-based LLR detector}, the following procedure is used to directly decode user B's signal:
	For direct decoding of user $\beta$'s signal, the \enquote{constellation rotation-based LLR detector} is employed as follows.
	\par \textit{1) First stage:} Identifying the active subcarrier indices is the receiver's first stage objective. To accomplish this, it first calculates a log-likelihood ratio for each subcarrier index, taking into account that the symbols in the frequency domain can have either non-zero or zero values. For index $\delta$, the first stage log-likelihood ratio can be expressed as follows:
	\begin{equation}
		\lambda _{stage1,\beta}(\delta )= \ln\frac{\sum_{\mu=0}^{M_{\beta}-1}P(x_{\beta}(\delta )=q_{\mu}^{\beta}\mid y_{\beta}(\delta ))}{P(x_{\beta}(\delta )=0\mid y_{\beta}(\delta ))}
		\label{Rotated LLR1}.	
	\end{equation}
 Further~\eqref{Rotated LLR1} can be written \cite{6587554} as follows by applying the Bayes' formula: 
	%\lambda _{stage1,\beta}(\delta )= \ln\frac{\sum_{\nu=0}^{M_{\beta}-1}P(x_{\beta}(\delta )=q_{\nu}^{\beta}| y_{\beta}(\delta ))}{P(x_{\beta}(\delta )=0\mid \overline y_{\beta}(\delta ))}

\begin{multline}
	\lambda _{stage1,\beta}(\delta )= \ln({k_{\beta}})-\ln({n_{\beta}-k_{\beta}})+\frac{\left|y_{\beta}(\delta ) \right|^{2}}{\sigma_{N_{F}}^{2}}\\
	+\ln\left (\sum_{\mu=0}^{M_{\beta}-1}\exp\left (\frac{-\left|y_{\beta
		}(\delta )-\sqrt{P_{\beta}}h_{F,\beta}(\delta )q_{\mu}^{\beta} \right|^{2}}{\sigma_{N_{F}}^{2}}\right )\right )
		\label{Rotated LLR2},
\end{multline}
	where $q_{\mu}^{\beta} \in {Q}'_{M_{\beta}}$ and $\delta=1, \ldots, N$. Here  $\beta\in\{A,B\}$.\\

~\eqref{Rotated LLR2} can be simplified \cite{6587554} as follows for the BPSK modulation:
	
	\begin{multline}
		\lambda _{stage1,\beta}(\delta )=\max(c_{\beta},d_{\beta})+\frac{\left|y_{\beta}(\delta ) \right|^{2}}{\sigma_{N_{F}}^{2}}\\
		+\ln\left (1+\exp\left (-\left|d_{\beta}-c_{\beta} \right|\right ) \right ) 
			\label{Rotated LLR3},
	\end{multline}
	
	where $c_{\beta}=-\left|y_{\beta}(\delta )-\sqrt{P_{\beta}}h_{F,\beta}(\delta )q_{1}^{\beta}  \right|^{2}/\sigma_{N_{F}}^{2}$ and $d_{\beta}=-\left|y_{\beta}(\delta )-\sqrt{P_{\beta}}h_{F,\beta}(\delta )q_{0}^{\beta}  \right|^{2}/\sigma_{N_{F}}^{2}$. From~\eqref{Rotated LLR3}, $N$ number of LLR values are determined. Then, for each subblock $g$, we compute a total $R_{\beta}$ number of LLR sums for the $R_{\beta}$ sets of feasible active indices combinations provided by the associated look-up table, where $R_{\beta }=2^{p_{\beta,1}}$. The set containing all feasible active indices combinations for the  $g$ th subblock is indicated as
	   $\psi_{g, \beta}=\left\{I_{g, \beta}^{1}, \ldots, I_{g, \beta}^{R_{\beta}}\right\}$, where $I_{g, \beta}^{\nu}=\left\{i_{g, \beta, 1}^{\nu}, \ldots, i_{g, \beta, k_{\beta}}^{\nu}\right\}$ for $\nu=1, \ldots, R_{\beta}$. The expression $J_{g ,\beta}^{\nu }=\sum_{\zeta =1}^{k_{\beta}}\lambda _{stage1,\beta}\left ( n_{\beta}(g -1)+i_{g ,\beta,\zeta }^{\nu }\right )$ computes the LLR sums that correspond to the $\nu$ th active subcarrier indices set, where $g=1, \ldots, G_{\beta}$ and $\nu=1, \ldots, R_{\beta}$. Thus, for each subblock g, a total $R_{\beta}$ number of LLR sums are obtained. The receiver selects the active indices set with the highest LLR sum among them. That is $\widehat{\nu }_{g }=\arg\max\limits_{\nu}J_{g ,\beta}^{\nu }$ and the detected active indices set is $I_{g, \beta}^{\widehat{\nu}_{g}}$.
	   
	   \par \textit{2) Second stage: } Finding constellation symbols corresponding to the detected active subcarrier indices set is the receiver's second stage objective. In order to do that, it calculates a log-likelihood ratio for each subcarrier index, taking into account that the frequency domain symbols can have a value of either $q_{0}^{\beta}$ or $q_{1}^{\beta}$, under the assumption that the modulation type is BPSK modulation. for index $\delta$, it is possible to compute this second stage log-likelihood ratio as follows. 
	\begin{equation}
		\lambda _{stage2,\beta}(\delta)=\ln\frac{P\left (  x_{\beta}(\delta )=q_{0}^{\beta}\hspace{0.5 mm}|\hspace{0.5 mm}y_{\beta}(\delta )\right )}{P\left (  x_{\beta}(\delta )=q_{1}^{\beta}\hspace{0.5 mm}|\hspace{0.5 mm}y_{\beta}(\delta )\right )}.
		\label{rotated stage2 LLR}		
	\end{equation}
	Applying Bayes' formula,~\eqref{rotated stage2 LLR} is expressed as:
	\begin{equation}
		\lambda_{stage2,\beta}(\delta)=d_{\beta}-c_{\beta} , \quad\delta=1, \ldots, N.
	\end{equation}
The second stage receives knowledge from the first stage regarding the detected active indices set. This set's denotation for the g th subblock is $I_{g, \beta}^{\widehat{\nu}_{g}}=\left\{i_{g, \beta, 1}^{\widehat{\nu}_{g}}, \ldots, i_{g, \beta, k_{\beta}}^{\widehat{\nu}_{g}}\right\}$, where $i_{g, \beta, \zeta}^{\widehat{\nu}_{g}} \in\left\{1, \ldots, n_{\beta}\right\}$ for $\zeta=1, \ldots, k_{\beta}$, here  $\widehat{\nu}_{g} \in\left\{1, \ldots, R_{\beta}\right\}$ for $g=1, \ldots, G_{\beta}$. The following is the algorithm of the second stage symbol decoding:

\begin{algorithm}
	\caption{Second stage decoding algorithm}
	\begin{algorithmic}[1]
		\FOR {$\beta= A,B $}    
		
		\FOR {$g= 1$ to $G_{\beta}$}
		  
		\FOR {$\zeta= 1$ to $k_{\beta}$}
		   
		\IF {$\lambda_{stage2,\beta}^{g}\left(i_{g, \beta, \zeta}^{\widehat{\nu}_{g}}\right) \geq 0$}
		
		\STATE decoded symbol is $q_{0}^{\beta}$ i.e. $\overrightarrow{x}_{\beta}^{g}\left(i_{g, \beta, \zeta}^{\widehat{\nu}_{g}}\right)=q_{0}^{\beta}$;
		   	
		\ELSIF{$\lambda_{stage2,\beta}^{g}\left(i_{g, \beta, \zeta}^{\widehat{\nu}_{g}}\right)<0$}
		
		\STATE decoded symbol is $q_{1}^{\beta}$ i.e.
		$\overrightarrow{x}_{\beta}^{g}\left(i_{g, \beta, \zeta}^{\widehat{\nu}_{g}}\right)=q_{1}^{\beta}$;
		   	
		\ENDIF
		   
		\ENDFOR
		\ENDFOR
	    \ENDFOR
		   
	   \end{algorithmic}
 \end{algorithm}
	
  Here $\lambda_{stage2, \beta}^{g}\left(i_{g,\beta,\zeta}^{\widehat{\nu}_{g}}\right)=\lambda_{stage2, \beta}\left(n_{\beta}(g-1)+i_{g,\beta,\zeta}^{\widehat{\nu}_{g}}\right)$. Following the preceding steps, the decoded $g$ th subblock of user $\beta's$ signal is formed as $\overrightarrow{\boldsymbol{x}}_{\beta}^{g}=\left[\overrightarrow{x}_{\beta}^{g}(1) \ldots \overrightarrow{x}_{\beta}^{g}\left(n_{\beta}\right)\right]^{T} $, where $\overrightarrow{x}_{\beta}^{g}(\varphi) \in\left\{0, q_{0}^{\beta}, q_{1}^{\beta}\right\}$ for $\varphi=1, \ldots, n_{\beta}$.\\

	\section{PERFORMANCE ANALYSIS}
In this section, the computational complexity of the proposed detector is investigated at both user A and user B, and it is compared with the existing detectors in the literature. Computational complexity is quantified via the number of complex multiplications.
\par At user B, the computational complexity of the \enquote{ML detector} derived from [8, eq.(6)] is $\sim{\cal O}\left(k_{B}R_{B}\left(M_{B}\right)^{k_{B}}\right)$ per subblock, and the computational complexity of the \enquote{rotated constellation based ML detector} derived from~\eqref{rotated ML} is $\sim{\cal O}\left(k_{B}R_{B}\left(M_{B}\right)^{k_{B}}\right)$ per subblock. That is, the \enquote{ML detector} and the \enquote{rotated constellation based ML detector} have the same computational complexity at user B. On the other hand,
The computational complexity of the \enquote{two-stage LLR detector} obtained from [14, eq.(5)] is $\sim{\cal O}\left(n_{B} M_{B}\right)$ per subblock, and the computational complexity of the \enquote{proposed detector} obtained from~\eqref{Rotated LLR2} is $\sim{\cal O}\left(n_{B} M_{B}\right)$ per subblock. In other words, the computational complexity of the \enquote{two-stage LLR detector} and the \enquote{proposed detector} comes out to be the same at user B. Note that, for user B, $\beta=B$ in~\eqref{rotated ML} and~\eqref{Rotated LLR2}.

%the two-stage LLR detector and the proposed detector have the same computational complexity at user B. Note that, for user B, $\beta=B$ in (1) and (3).
  \par At user A, the total computational complexity of the \enquote{ML detector} computed from [8, eq.(3)] and [8, eq.(5)] is  $\sim{\cal O}\left(k_{B} R_{B}\left(M_{B}\right)^{k_{B}}+\right.$ $\left.k_{A} R_{A}\left(M_{A}\right)^{k_{A}}\right)$ per subblock. While the computational complexity of the \enquote{rotated constellation-based ML detector} calculated from~\eqref{rotated ML} is $\sim{\cal O}\left(k_{A}R_{A}\left(M_{A}\right)^{k_{A}}\right)$ per subblock. On the other hand, at user A, the total computational complexity of the \enquote{two-stage LLR detector} determined from [14, eq.(10)] and [14, eq.(15)] is  $\sim{\cal O}\left(n_{B} M_{B}+n_{A} M_{A}\right)$ per subblock. while the computational complexity of the \enquote{proposed detector} derived from ~\eqref{Rotated LLR2} is $\sim{\cal O}\left(n_{A} M_{A}\right)$ per subblock. Note that, for user A, $\beta=A$ in~\eqref{rotated ML} and~\eqref{Rotated LLR2}.
   \par From the above analysis, it can be seen that the proposed detector has significantly less computational complexity than the conventional ML detector. At user B, the reason for the much reduced complexity of the proposed detector is the log-likelihood ratio (LLR) based algorithm used in it. While at user A, there are two reasons for the complexity reduction of the proposed detector. First, due to the use of the LLR-based algorithm. Second, due to the use of the rotated constellation-based concept, the receiver can directly decode user A's signal without needing to go through the SIC process at user A.  For better understanding, the following TABLE I, TABLE II, and TABLE III show the complexity reduction of different detectors as compared to the conventional ML detector in various scenarios.
	\begin{table}[!h]
		\caption{Complexity reduction as compared to the optimal ML detector where both the users having mid data rate application (i.e. $k_{A}=k_{B}=2$, $n_{A}=n_{B}=n=4$, $M_{A}=M_{B}=M=2$).}
		\begin{center}
			\resizebox{.47\textwidth}{!}{
			\begin{tabular}{|c|c|c|}
				\hline
				%\multicolumn{6}{c}{} \\
				%\hline
	\textbf{Detector type} & \textbf{Complexity reduction at user B} & \textbf{Complexity reduction at user A} \\
				
				%\hline
				%\multicolumn{4}{c}{Encoder} \\
				\hline
				\hline
				Rotated ML & 0\% &  50\% \\
			
				\hline
				
				Two-stage LLR & 75\% &  75\%\\
			
				\hline
			
				Proposed & 75\% &  87.5\%\\
			
				\hline
			\end{tabular}}
		\end{center}
		
		\label{mid table}
	\end{table}

		\begin{table}[!h]
		\caption{Complexity reduction as compared to the optimal ML detector where both the users having high data rate application (i.e. $k_{A}=k_{B}=3$, $n_{A}=n_{B}=n=4$, $M_{A}=M_{B}=M=2$).}
		\begin{center}
			\resizebox{.47\textwidth}{!}{
			\begin{tabular}{|c|c|c|}
				\hline
				%\multicolumn{6}{c}{} \\
				%\hline
			\textbf{Detector type} & \textbf{Complexity reduction at user B} & \textbf{Complexity reduction at user A} \\
			
				%\hline
				%\multicolumn{4}{c}{Encoder} \\
				\hline
			    \hline
				
				Rotated ML & 0\% &  50\% \\
			
				\hline
				
				Two-stage LLR & 91.7\% &  91.7\%\\
				
				\hline
			
				Proposed & 91.7\% &  95.8\%\\
			
				\hline
			\end{tabular}}
		\end{center}
		
		\label{high table}
	\end{table}
	
 \begin{table}[!h]
		\caption{Complexity reduction as compared to the optimal ML detector for the hybrid user specification (i.e. $k_{A}=3, k_{B}=1$, $n_{A}=n_{B}=n=4$, $M_{A}=M_{B}=M=2$).}
		\begin{center}
			\resizebox{.47\textwidth}{!}{
			\begin{tabular}{|c|c|c|}
				\hline
				%\multicolumn{6}{c}{} \\
				%\hline
			\textbf{Detector type} & \textbf{Complexity reduction at user B} & \textbf{Complexity reduction at user A} \\
			
				%\hline
				%\multicolumn{4}{c}{Encoder} \\
				\hline
			    \hline
			Rotated ML & 0\% &  25\% \\
				
				\hline
			
				Two-stage LLR & 75\% &  87.5\%\\
			
				\hline
			
				Proposed & 75\% &  93.75\%\\
			
				\hline
			\end{tabular}}
		\end{center}
		
		\label{hybrid table}
	\end{table}

	\section{SIMULATION RESULTS}
	\begin{figure}[h!]
		\centering
        \includegraphics[width=\linewidth]{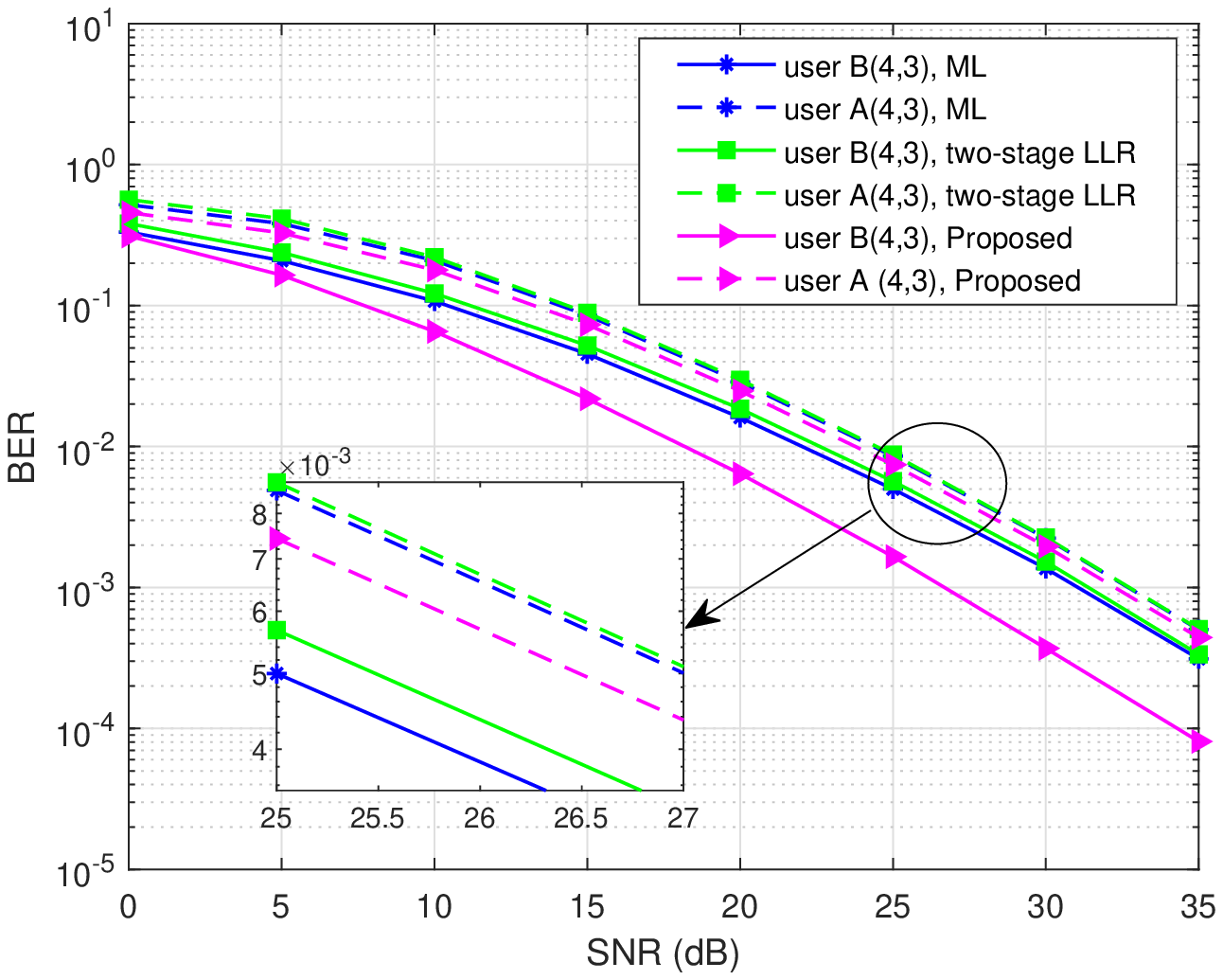}
		\caption{Comparing BER of the proposed detector with the \enquote{ML detector} and the \enquote{two-stage LLR detector}, where both users have high data rate applications}
		\label{High Use}
	\end{figure}
	
	\begin{figure}[ht!]
	\centering
	\includegraphics[width=\linewidth]{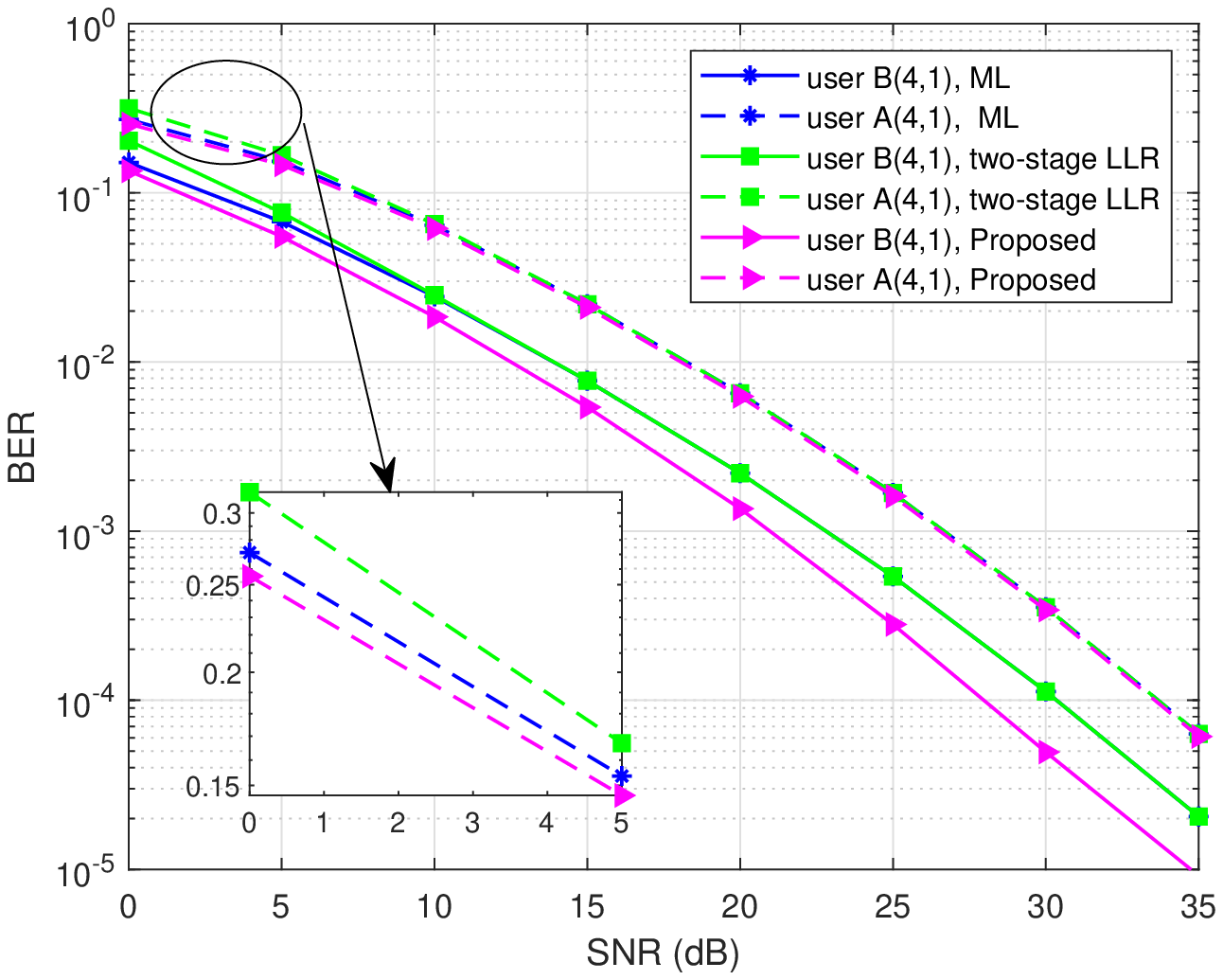}
	\caption{Comparing BER of the proposed detector with the \enquote{ML detector} and the \enquote{two-stage LLR detector}, where both users have low data rate applications}
	\label{Low Use}
   \end{figure}

\begin{figure}[ht!]
	\centering
	\includegraphics[width=\linewidth]{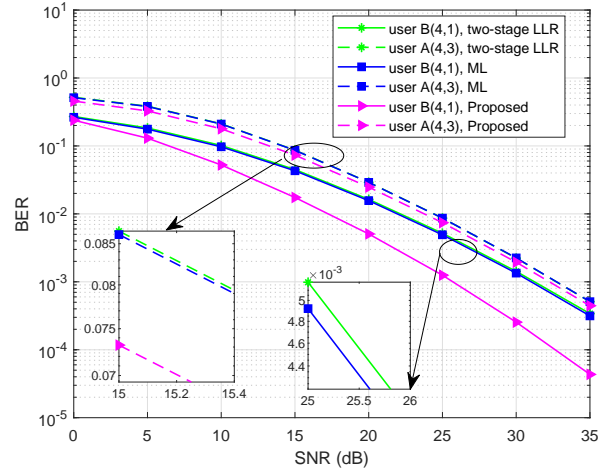}
	\caption{Comparing BER of the proposed detector with the \enquote{ML detector} and the \enquote{two-stage LLR detector} for the hybrid user configuration}
	\label{Hybrid Use}
\end{figure}

\begin{figure}[ht!]
	\centering
	\includegraphics[width=\linewidth]{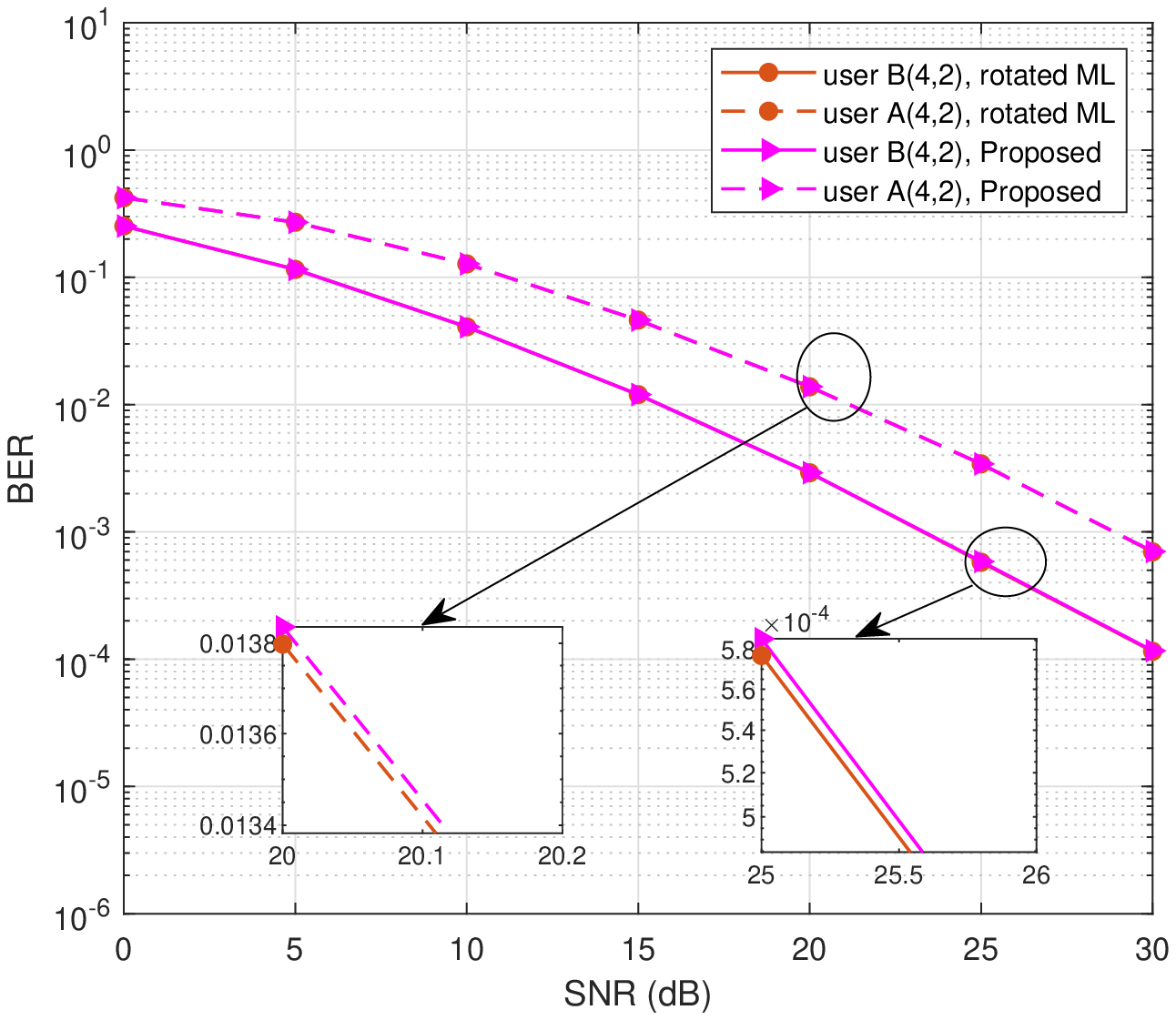}
	\caption{Comparing BER of the proposed detector with the BER of the \enquote{rotated constellation-based ML detector}}
	\label{rotated ml vs llr}
\end{figure}

This section presents the simulation results of the error performance of the proposed detector in presence of the Rayleigh fading channel and compares it with the error performance of the \enquote{ML detector}, the \enquote{two-stage LLR detector} \cite{9844730} and the \enquote{rotated constellation based ML detector}. The system parameters used for all the simulations are: $N=256, G_{A}=G_{B}=64, n_{A}=n_{B}=n=4$, $\varrho_{A}^{2}=$$4 \mathrm{~dB}, \varrho_{B}^{2}=0 \mathrm{~dB}, N_{cp}=16, L=12, \gamma=0.1$, and $M_{A}=M_{B}=M=2$.
	\par With a same-user application taken into account, Fig.~\ref{High Use} compares the BER performance of the proposed detector with the BER performance of the \enquote{ML detector} and the \enquote{two-stage LLR detector}. By setting the simulation parameter to $k_{A}=k_{B}=3,n=4$, it is assumed that both users having high data rate applications (e.g. ultrahigh definition video streaming) utilize the same NOMA spectrum. Therefore, an identical subcarrier activation ratio of 3/4 is considered for both users in this case. From this figure, we can observe that at user B, the proposed detector achieves a gain of 4 dB at a BER of $10^{-3}$, when compared to the \enquote{ML detector} and the \enquote{two-stage LLR detector}. While, at user A, the proposed detector attains a slightly better error performance than the optimal ML detector and the two-stage LLR detector at a BER of $10^{-3}$.
	
	\par Considering another same-user application, Fig.~\ref{Low Use} compares the BER performance of the proposed detector with the BER performance of the \enquote{ML detector} and the \enquote{two-stage LLR detector}. By setting the simulation parameter to $k_{A}=k_{B}=1,n=4$, it is assumed that both users having low data rate applications (e.g. Internet of Things application) are utilizing the same NOMA spectrum. Therefore, an identical subcarrier activation ratio of 1/4 is considered for both users in this case. According to this figure, at user B, the proposed detector achieves a gain of 2 dB at a BER of $10^{-3}$, when compared to the \enquote{ML detector} and the \enquote{two-stage LLR detector}. While at user A, the proposed detector achieves a minute improvement in error performance over the ML detector and the two-stage LLR detector at a BER of $10^{-3}$.
 
 %It can be seen from Fig.1 that the error performance for the proposed low complexity detector is almost the same as the error performance of the ML-based detector.
	\par For a hybrid-user application, Fig.~\ref{Hybrid Use} compares the BER performance of the proposed detector with the BER performance of the \enquote{ML detector} and the \enquote{two-stage LLR detector}. Since user A and user B with different settings are utilizing the same spectrum, the users in this scenario are referred to as hybrid users. By setting the simulation parameter to $k_{A}=3, k_{B}=1, n=4$, it is considered that the user A with a high data rate application and user B with a low data rate application utilize the same NOMA spectrum. Therefore, a subcarrier activation ratio of 3/4 and 1/4 is assigned to user A and user B, respectively. Fig.~\ref{Hybrid Use} shows that at user B, the proposed detector achieves a gain of 5.5 dB at a BER of $10^{-3}$ when compared to the \enquote{ML detector} and the \enquote{two-stage LLR detector}. While, at user A, the proposed detector achieves a slightly better error performance than the optimal ML detector and the two-stage LLR detector at a BER of $10^{-3}$.

	\par In Fig.~\ref{rotated ml vs llr}, the error performance of the proposed detector is compared with that of the \enquote{rotated constellation-based ML detector}. This figure illustrates that the proposed detector obtains nearly identical error performance as the \enquote{rotated constellation-based ML detector}, while the proposed detector has much reduced computational complexity than the \enquote{rotated constellation-based ML detector} at both user A and user B, as indicated in the previous performance analysis section.
	\section{CONCLUSION}
	In this paper, a novel \enquote{constellation rotation-based LLR detector} is proposed for the OFDM-IM NOMA scheme. This proposed detector is evaluated under several scenarios, including users with high data rate applications, low data rate applications, and hybrid applications. Simulation results and complexity analysis demonstrate that in every case, the proposed detector achieves significantly lower computational complexity while also achieving much better error performance than the earlier introduced detectors in the literature.
	
	%\bibliography{REF_TWO_STAGE_LLR}
	\bibliographystyle{IEEEtran}
	%\bibliography{IEEEabrv,./REFBib}
	%\bibliographystyle{plain}
	\bibliography{ROTATED_LLR_WHOLE}

	%\printbibliography
	
\end{document}